# The Sower's Way: Quantifying the Narrowing Net-Energy Pathways to a Global Energy Transition

**Authors:** Sgouris Sgouridis[1]*, Ugo Bardi[2], Denes Csala[1]

**Affiliations:**

[1] Institute Center for Smart and Sustainable Systems, Masdar Institute of Science and Technology. United Arab Emirates. Tel: +971 2 810 9156, Fax: +971 2 810 9901.

[2] University of Florence, Department of Earth Sciences, C/o Dipartimento di Chimica, Polo Scientifico di Sesto Fiorentino, 50019, Florence, Italy. Email: ugo.bardi@unifi.it

*Correspondence to: ssgouridis@masdar.ac.ae

**Abstract**:

Planning the appropriate renewable energy installation rate should balance two partially contradictory objectives: substituting fossil fuels fast enough to stave-off the worst consequences of climate change while maintaining a sufficient net energy flow to support the world's economy. The upfront energy invested in constructing a renewable energy infrastructure subtracts from the net energy available for societal energy needs, a fact typically neglected in energy projections. Modeling feasible energy transition pathways to provide different net energy levels we find that they are critically dependent on the fossil fuel emissions cap and phase-out profile and on the characteristic energy return on energy invested of the renewable energy technologies. The easiest pathway requires installation of renewable energy plants to accelerate from 0.12TW$_p$/year in 2013 to peak between 6.6 and 10.4 TW$_p$/year, for an early or a late fossil-fuel phase-out respectively in order for emissions to stay within the recommended CO$_2$ budget.

Note: due to size limits, the Supplementary Information cannot be uploaded to Arxiv. Please contact us (ssgouridis@masdar.ac.ae) and we can share the file directly.

**Background**

The 21st Conference of the Parties (COP21) has highlighted the need for a rapid transition away from fossil fuels in order to maintain the emissions of greenhouse gases below a level considered to have an acceptable probability of being safe, leading to a temperature increase not greater than 2C or even 1.5C. While the goal is clear and ambitious, the methods for attaining it are not. The Intended Nationally Determined Contributions (INDCs) submitted prior to the conference are clearly insufficient to maintain emissions under the target level and the current mainstream scenarios presented rely on, controversial, late century negative emissions [1],[2]. In addition to the need of reducing greenhouse emissions, a second fundamental target that is implicit in the COP21 agreement is that these reductions should be obtained while offering sufficient available energy for humankind, especially for developing countries that are ascending the energy availability ladder.

Such a transition implies the development of a new energy paradigm, relying on renewable energy (RE) and supported by a more efficient infrastructure for its storage, transmission and use. Is it possible to satisfy the dual constraint of reducing emissions fast enough while achieving the desired energy availability? Posed this way, it becomes a physical rather than an economic problem. In order to build the new infrastructure and power the transition, we need energy that, initially, will primarily come from fossil fuels. As the transition progresses, fossil fuels are replaced by RE to the point that the energy system relies 100% on the latter. The transition is therefore determined by the rate of RE installations that substitutes the fossil removals but also by the energy investment needed to support this rate as defined by the energy return on energy investment (EROEI) of the sources. The economic approaches employed by the vast majority of energy models used for long-term planning fails to account for this crucial detail while relying on negative emissions. We present a methodology for modeling the dynamics of the energy transition using the more appropriate net energy approach.

While conventional energy statistics track gross energy, the crucial metric for sustaining socio-economic metabolism is *net energy*, the energy made available from a resource after subtracting the energy expended in its extraction, upgrade and distribution [3],[4]. When the EROEI [5] is large and the investment is operational i.e. concurrent with the energy extraction, like in fossil fuels and biomass, the net or gross distinction is of limited consequence. It becomes critical though when EROEI is lower, the investment is capital intensive, as in the case of technical renewable energy (RE) like wind or solar, and the rate of installation is high [6]. Accounting for these dynamic effects, we map the possible trajectories for a sustainable energy transition (SET) at a global scale.

We show that renewable energy installation rates should accelerate and increase at least by a factor of 50 and perhaps more than 90 over current, at their peak, to mitigate climate change and sustain the economy. These estimates are based on a global energy model that uses physical energy balances to estimate net, as opposed to gross, energy availability. It is the first model to transparently represent the dynamics of the energy transition on a physical rather than on an economic basis. With the reasoned assumption that options like efficiency, carbon capture, nuclear and biomass that were central in previous work cannot scale sufficiently fast, we project specific annual renewable energy installation rates that form trajectories compliant with a rigorously defined SET.

A key characteristic of SET is that it requires energy to construct the necessary RE infrastructure and to integrate the mostly variable RE resources in the energy system. Since at present the world's energy derives primarily from fossil resources, *we need energy from fossil fuels to transition away from their use*. This requirement is analogous to "the sower's strategy" [7], the long-established farming practice to save a fraction of the current year's harvest as seeds for the next. Fossil fuels produce no "seed" of their own but we can "sow" what these fuels provide: energy and minerals to create the capital needed for the transition [8]. Yet, withdrawing the "seed" energy reduces net available energy for society. The challenge therefore is to balance energy availability and emissions in order to complete a renewable transition before fossil fuel depletion makes it impossible without inflicting crippling damages on the climate. Past transitions have been partial and yet slow, typically spanning several decades [9]. However, a SET should encompass the entire energy system while providing sufficient *net* energy to sustain the global economy. We should also account for the decreasing quality of the fossil fuel resources, as a depleting stock requires additional energy for extraction and refining [10].

Since economic output requires energy availability [11] (also cf. Fig. SI1), currently reliant on carbon-intensive fuels, climate mitigation is perceived as a trade-off between climate and energy. Rather than carbon emissions, we refocus on providing sufficient RE for the global energy needs at a rate that will make fossil fuel combustion redundant, and thus leave unburned a fraction of the fossil carbon reserve [12]. From this perspective, the rate of renewable energy installations becomes the defining parameter of a SET trajectory.

Previously, the "wedges" approach identified discrete efficiency and energy measures that produce equivalent emissions reductions and stacked them cumulatively [13,14]. While useful in breaking down the emissions problem and making it tractable, it did not account for the depletion profile of fossil fuels and overlooked the energy investment for constructing the wedges that is subtracted from the gross energy flows becoming unavailable for other productive uses. Similarly ignoring net energy impacts, forecasts based on economics also tend to be biased by contemporary conditions and misjudge "surprises" [15]. This materializes as a consistent underestimation of the growth potential of renewable energy (RE) in forecasts. Work on low representative concentration pathways (RCP), used as basis for the 2°C compliant Intergovernmental Panel on Climate Change (IPCC) scenarios, overwhelmingly relied on biomass, nuclear, and carbon capture and storage (CCS) along with energy efficiency [16] overstating the potential of these options to scale and underestimating their relative costs against RE.

**Defining and Modeling SET**
So what is the appropriate rate and schedule to make the energy transition sustainable? We define SET using three normative statements drawing from ecological economics [17], [18] (discussed further in the Supplementary Information, SI Section 1):
  I. *the impacts from energy use during SET should not exceed the long-run ecosystem carrying and assimilation capacity*
  II. *per capita net available energy should remain above a level that satisfies societal needs at any point during SET and without disruptive discontinuities in its rate of change*

III. *the rate of investment in building renewable energy harvesting and utilization capital stock should be sufficient to create a sustainable energy supply basis without exhausting the non-renewable safely recoverable resources.*

SET, therefore, depends on three crucial parameters: the rate at which fossil fuels can be safely combusted, the net energy society requires to function, and the characteristic EROEI of the substitute RE infrastructure portfolio. Matching net energy demand to an energy supply system that transitions from fossil fuels to RE forms a dynamic problem. Current RE investments subtract from the available energy today *and* shape future energy availability. To address this, we develop a physical energy-balance model that bridges the constraints articulated by the first two sustainability statements with the stock-flow dynamics of the third in order to map the space of SET-compliant energy system trajectories (SI Section 2).

The IPCC reports a probabilistic carbon emissions cap with uncertainty stemming from the different climatic models - we investigate the implications of the three levels that restrict warming to below 2C (510, 990, 1505 Gt $CO_2$) [19]. These provide a cumulative limit which can, in turn, be met by the different phase-out strategies shown in Fig. SI2: an early peak and gentler phase-out slope, a substitution of more carbon-intensive fossil fuels (e.g. coal) with less intensive ones (e.g. natural gas), or a delayed peak that forces a very steep phase-out afterwards.

Similarly, future energy demand depends on assumptions for the energy intensity of the economy and the infrastructure needed to store and use RE, but also on the level of convergence between developed and developing economies. In terms of gross average power per capita, estimates range from a low 1400W [20], 2000W as a lower limit for a high-income society [21], up to 10000W [5,22]. A study of a deep, yet still partial decarbonization, for the state of California modeled the transition from 6570W/person in 2010 to 3800W/person by 2050 with aggressive energy efficiency but without accounting for energy embodied in imported goods [23]. Figure SI3 shows the range of potential demand trajectories we investigate.

The Results section present a detailed transition trajectory for one set of assumptions on net primary energy requirements, fossil fuel phase-out strategy, and the RE portfolio EROEI. Given the inherent uncertainties, this trajectory is but a cross-section of a larger transition landscape. We investigate and map this landscape of SET trajectories for a range of: demand, EROEI, fossil-fuel phase-out strategies (early peak, switch to natural gas, and late peak), as they all significantly impact the required RE installation rates. Since some trajectories are more taxing than others, we calculate a feasibility index for each trajectory showing that the more feasible SETs form a thin band that gets narrower as we delay the fossil-fuel phase-out.

**Results**

The primary energy supply side of a SET trajectory for a given net final demand is fully described by RE installed capacity and the annual RE installation rate. The latter presents a simple but defining parameter of the transition that serves as a clear measure for planning purposes as it can show effectively how closely actual rates track the desired trajectory. Fig. 1 shows an example of a possible SET trajectory that presents the details of all constitutive energy resources for a 2000W net energy per capita demand by 2100 and an initial weighted-average RE EROEI of 20. The RE energy investment magnitude (the difference between gross and net energy) is evident as the notable hump above the dashed line in Fig. 1a during the transition acceleration phase (2020-2060), highlighting the role of fossil fuels as "seed" of the transition. Fig. 1b and 1c respectively show the RE installed capacity and installation rate.

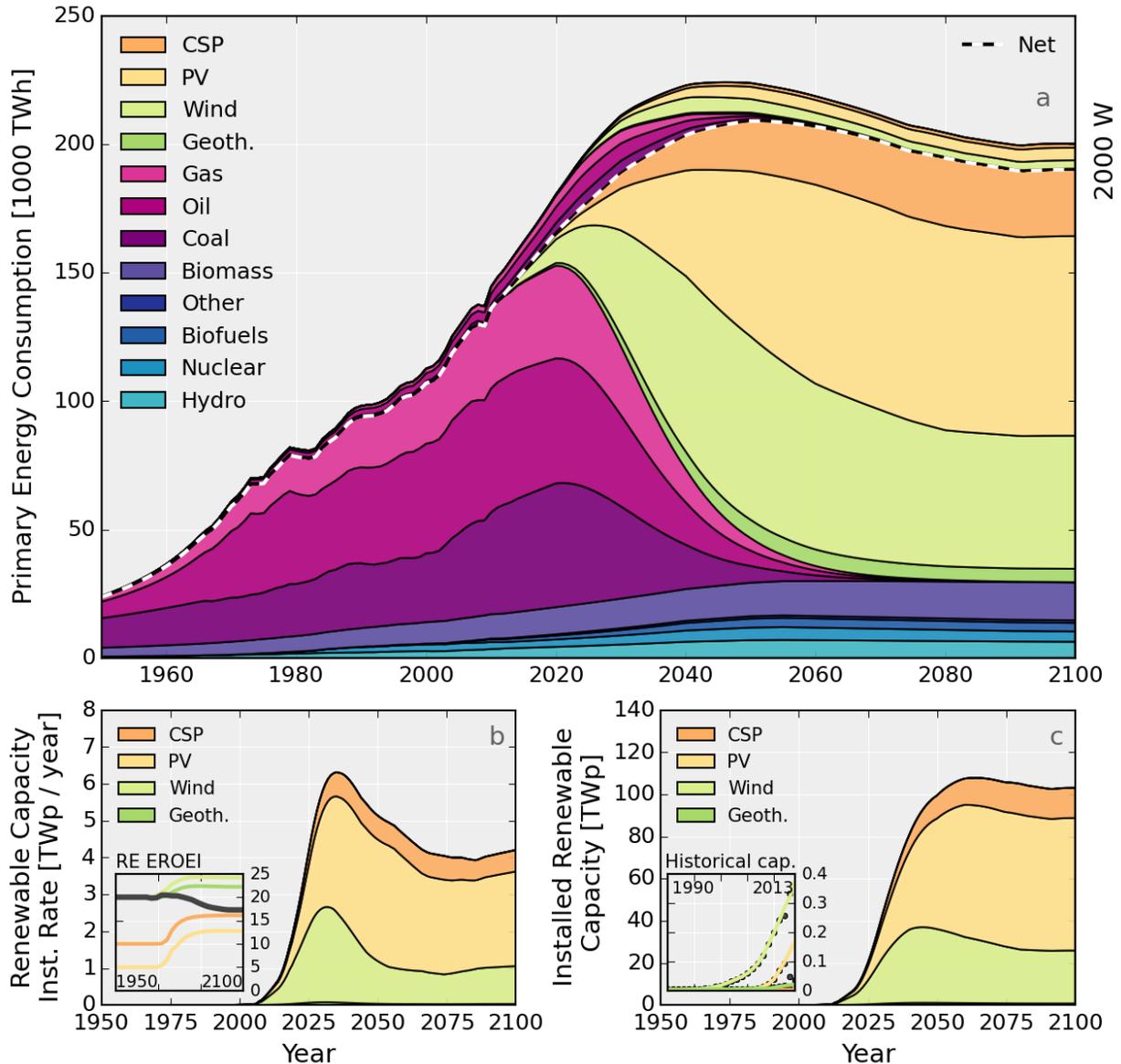

**Fig. 1**: (a) SET-compliant primary energy supply evolution (in PWh) for providing 2000W average net power per capita by 2100 to a population of 10.8 billion. Fossil fuel emissions comply with a 990 Gt $CO_2$ cap peaking in 2020 and phased-out by 2075. (b) RE portfolio installation rate profile (in $TW_p$/year). Inset shows the evolution of the weighted-average, composite RE EROEI (black line) with an initial value of 20 and the EROEI values for each technology. (c) Installed RE Capacity (in $TW_p$). Inset magnifies the 1990-2014 historical values (dotted lines) versus the modeled curves.

Given the wide possible range in both the EROEI of the RE supply and the net demand for energy, we compile the SET-compliant trajectories of RE installation rates and capacity into contour maps to illustrate the impact of these parameters on the RE trajectories. The isolines of

Fig. SI4 and SI5 show the required RE installed capacity for the range of power demand profiles in Fig. SI3b and a range of RE EROEI (6.7-60) respectively under the three fossil fuel caps and the three fossil phase-out schemes. Each horizontal cross-section represents a SET-compliant RE trajectory and the slope of the contours represents the net capacity addition rate. The actual installation rate, mapped in Figures SI6 and SI7, is larger as it accounts for the replacement of decommissioned RE installations.

Each mapped trajectory represents a combination of options that have different characteristic feasibility. For example, a higher per capita available energy makes it easier for society to prosper than a lower one and therefore is more desirable to sustain social cohesion [11]. Similarly, RE with higher EROEI values is more difficult to achieve than with lower ones as are trajectories that require investment of a higher fraction of the gross energy than those with lower. Finally, a transition that draws more energy from the gross available either in the form of a high peak or a high average is costlier to society in terms of resource expenditure. Since these four parameters are conflicting in their desirable range, we develop a multiplicative and normalized transition feasibility index (TFI) that weighs the relative difficulty" each assumption creates.

To make relative evaluation easier, Fig. 2 plots in profile the SET-compliant RE installation rates in $TW_p$/year for all combinations of EROEI and final demand trajectory colored by their relative TFI value. A lower EROEI pushes higher installation rates earlier by several years in response to the increased upfront investment. Nevertheless, in the critical initial acceleration phase, capacity additions are more influenced by the emissions cap and fossil phase-out. Looking at the highest TFI trajectory for the 990Gt $CO_2$ cap, the choice of an early fossil fuel phase-out halves the installation rates with corresponding peak at 6.6 in 2020 versus 10.4 TW/year for the late one in 2030 (comparing the bold lines in Fig. 2d and 2f). In either case, this implies an increase from the 2013 RE installations that were around 0.12TW/year by a factor ranging from 55 to 87. The early and fuel-switch phase out profiles offer a wider range of easier paths than the delayed transitions primarily because of their lower investment peaks. Critically, the lower 510Gt $CO_2$ cap leaves little slack creating a very narrow SET range penalized by the consistently high peak installation rates around $10TW_p$ although if the peak can be reached then the rest of the trajectory becomes achievable.

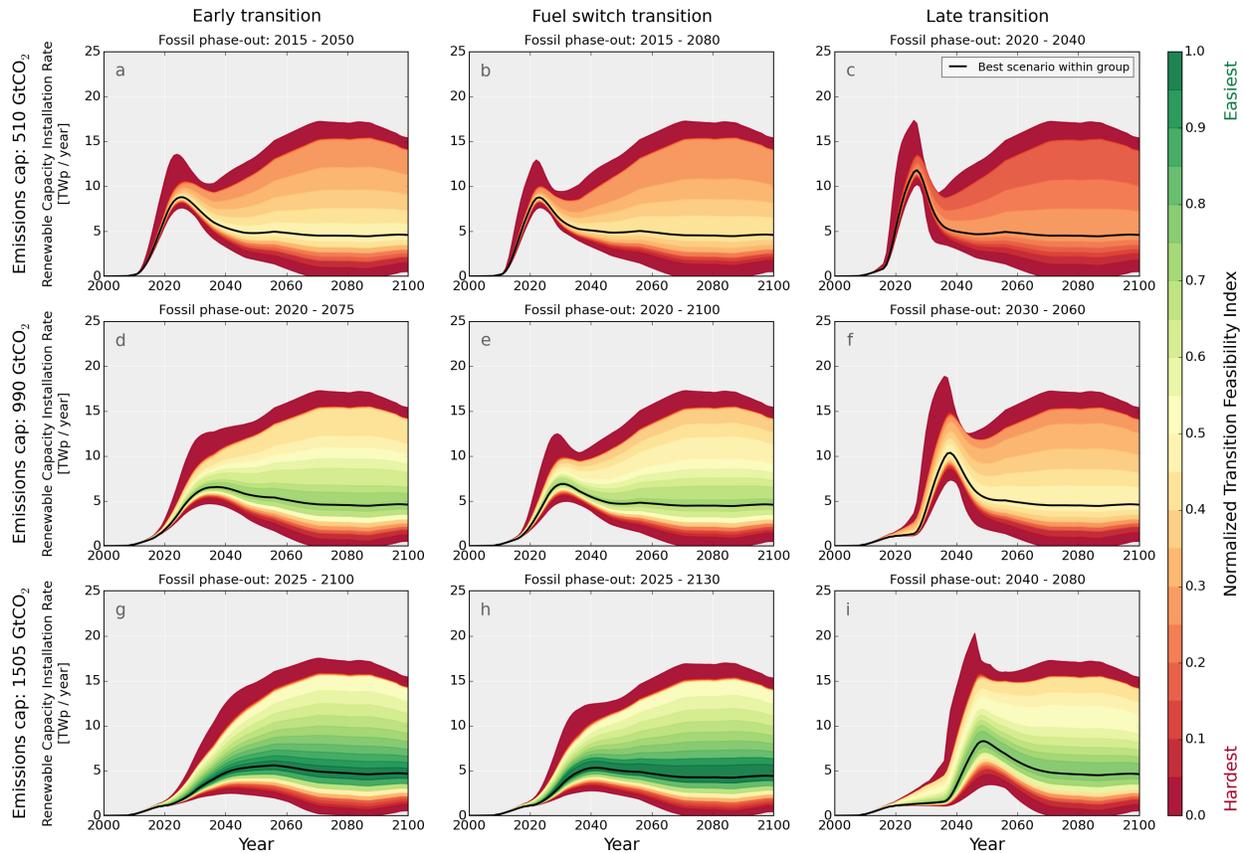

**Fig. 2**: Envelopes of SET-compliant RE installation rates over the RE composite EROEI range of 6.67 to 60 (in 2014) and a per capita net power demand range of 667-6000 W/person (in 2100), under early, early/partial and late fossil fuel phase-out strategies and for three fossil emission caps: 510 (a,b,c), 990 (d,e,f), and 1505 (g,h,i) Gt $CO_2$. The Transition Feasibility Index is a normalized composite measure of the relative difficulty of each trajectory. The solid lines draw the highest TFI (most feasible) trajectory in each set.

**Discussion**

In every case, a successful SET consists of a sustained acceleration in the rate of investment in renewable energy of more than one order of magnitude within the next three decades following a trajectory dictated by the chosen fossil-fuel phase-out. A peak in installation rates, but not cumulative capacity, forms at the point where the rate of energy demand growth starts to slow down. Transition trajectories that have a lower peak are harder as they assume an economy that can operate at the lowest energy per capita or very high EROEI values for RE. A fuel switch strategy helps in the second half of the century as it lowers the required installations but not so much at the acceleration phase. Importantly, further delays in the rise of RE investments cannot be compensated by subsequent additional acceleration because the decline in net energy from the constrained fossil fuels would be insufficient to power the transition without energetically impoverishing society or exceeding the carbon cap. Of course, attaining SET depends not only on RE installation but also on replacing a large part of the present energy-utilizing infrastructure (from industrial machines and vehicles to buildings and roads) to match the new energy resources.

While present infrastructure systems are adapted to specific energy carriers, for instance liquid fuels for transportation, the technologies available for SET provide electricity, a high quality energy carrier. The relatively low fossil fuel final energy conversion efficiency might imply that an RE-based, electrified economy would require much lower levels of per capita primary energy for the same economic output. In SET societies can adapt by means of restructuring their productive infrastructure and, if not possible, to use electricity to produce synthetic fuels. For instance, modern agriculture relies on fossil resources for mechanization and for the production of critical fertilizers. Nevertheless, it is technically possible to transform the agricultural infrastructure from one using liquid fuels to one that uses electricity [24]. In other cases, e.g. in aviation, synthetic drop-in fuels can be produced, but with large associated conversion losses. Results from network analysis indicate that future societal energy intensity is likely to be higher if expected urbanization trends materialize [25]. As a result, done on a global scale, an RE-based energy system will deploy (i) overcapacity (ii) mechanical, electro-chemical and chemical storage [26] for grid stability, daily and seasonal variability, and synthetic fuels, and (iii) replacement, modification and expansion of infrastructure systems to support rising incomes in the developing world and in order to make them compatible with renewable energy supply – e.g. electrification of transportation systems and long-distance electricity transmission. While our estimates provide average annual power they are conservative as either actual installed power may need to be higher to capture demand peaks and seasonal variation and/or storage systems will need to be deployed adding to the energy investment required for a functioning RE-based system. Given that there are significant lead times and many infrastructure investment decisions taken today have useful lives and impacts of several decades [27], their construction needs to account for SET already.

On the question of policy, after setting appropriate RE targets, the current economic and political system should generate the mechanisms necessary to prioritize the allocation of these resources from other activities. Economic activity appears to be a product of a reinforcing process between energy availability and expansion and therefore growth is reliant on a continuous increase in the quality-adjusted energy supply [28]. For a financial system in which debt is extended relying on the expectation that future growth will permit its repayment, this implies that it cannot stay solvent without securing an adequate energy supply to support the expected future economic energy intensity. We therefore propose a corollary, normative economic statement for SET on par with the physical ones: *financial commitments of future consumption (debt) should be limited by future energy availability*. Tying debt extension to RE investment could provide a self-regulating incentive to the financial system to pursue the energy transition.

**Conclusions**

The challenge of a sustainable energy transition before the end of the 21$^{st}$ century under climate constraints is unprecedented in magnitude, scope, and ambition. It is, nevertheless, doable if we adopt a global "sower's strategy" and invest an appropriate amount of the fossil energy available today into building a sustainable energy future with concrete annual targets. An energy metabolism perspective simplifies a notable confusion in the discussion of RE potentials [29] as it can provide a range for the RE investment effort (the "seed") and objectively inform policy formation by back-casting on the appropriate SET trajectory for a given desired net energy availability. The acceleration by a factor of 50 in RE installation rate is robust across trajectories in early and fuel switch transitions but delays in picking up pace may lead to rates that exceed current by a factor of 90 and more, making them rather impractical.

**Acknowledgments:**

We would like to thank Masdar Institute and University of Florence for supporting this research. Several colleagues have provided feedback to previous iterations of this text. We would like to specifically thank Michael Dale, Marco Raugei, Steve Griffiths, Scott Kennedy, Antonio Garcia-Olivares and Iyad Rahwan for their detailed comments.